\documentclass[fdp,fleqn]{w-art} 

\usepackage{times}
\usepackage{w-thm}
\usepackage{amsmath}
\usepackage{amscd}
\usepackage{amsfonts}
\usepackage{epsfig,multicol}
\usepackage{amsthm}
\usepackage[]{graphicx}
\def\IZ{\mathbb{Z}}
\def\IR{\mathbb{R}}
\def\ID{\mathbb{D}}

\newcommand\fverb{\setbox\pippobox=\hbox\bgroup\verb}
\newcommand\fverbdo{\egroup\medskip\noindent%
           \fbox{\unhbox\pippobox}\ }
\newcommand\fverbit{\egroup\item[\fbox{\unhbox\pippobox}]}
\newbox\pippobox

\begin{document}
\DOIsuffix{theDOIsuffix} 
\Volume{55}
\Issue{1}
\Month{01}
\Year{2007} 
\pagespan{1}{} 
\Receiveddate{2008} \Reviseddate{2008} \Accepteddate{2008}
\Dateposted{2008} 



\title[D-branes on generalized K\"{a}hler geometries: II. Dualities]{$\mathcal{N}$ = 2 world-sheet approach to D-branes on generalized K\"{a}hler geometries: II. Dualities}


\author[A. Sevrin]{Alexander Sevrin\inst{1,}%
  \footnote{E-mail: Alexandre.Sevrin@vub.ac.be}}
\address[\inst{1}]{Theoretische Natuurkunde, Vrije Universiteit Brussel and \\
The International Solvay Institutes,\\
Pleinlaan 2, B-1050 Brussels, Belgium}
\author[W. Staessens]{Wieland Staessens\inst{1,}\footnote {E-mail: Wieland.Staessens@vub.ac.be, FWO aspirant}}
\author[A. Wijns]{Alexander Wijns\inst{2,3,}\footnote{E-mail: awijns@nordita.org}}
\address[\inst{2}]{Department of Mathematics,
Science Institute, University of Iceland,\\
Dunhaga 2, 107 Reykjavik, Iceland.}
\address[\inst{3}]{NORDITA,
Roslagstullsbacken 23, 106 91 Stockholm, Sweden.}

\begin{abstract}
Following the general formalism reviewed in \cite{SSW:2009pr1} we
present several examples of possible D3-brane configurations on
four-dimensional generalized K\"{a}hler geometries. We will discuss
T-duality transformations in $\mathcal{N}$ = 2 boundary superspace
and apply the duality transformations to the constructed D3-branes.
The duality transformations lead to a systematic method to construct
coisotropic branes, even on target spaces that are not
hyper-K\"{a}hler.
\end{abstract}
\maketitle                   





\section{Motivational introduction} \label{sec1}
Already in the early days of supersymmetry it was realized that
there exists an intimate relation between extended supersymmetry and
complex geometry, when applied to (non-linear) $\sigma$-models
\cite{Zumino:1979, Alvarez-Gaume:1981hm, Gates:1984nk,
Buscher:1987uw, Sevrin:1996jr, Lindstrom:2005zr}. From a more modern
perspective - inspired by flux compactifications - the
two-dimensional $\mathcal{N}$ = (2,2) $\sigma$-model is part of the
stringy toolbox to study type II superstrings on internal manifolds
without R-R fluxes. The geometrical data to describe this type of
manifold is given by a metric, a closed 3-form and two complex
structures. The complex structures are covariantly constant and the
metric is hermitian with respect to both complex structures. This
type of geometry was originally called bihermitian geometry
\cite{Gates:1984nk}, but since the birth of generalized complex
geometry \cite{Gualtieri:2007ng} it is usually referred to as
generalized K\"{a}hler geometry. The conditions on the metric, the
3-form and the complex structures can be solved in terms of a single
real potential, the so-called generalized K\"{a}hler potential,
from which we can determine the metric and 3-form.\\
\indent In order to make the $\mathcal{N}$ = (2,2) manifest, to
simplify the analysis and to expose the geometrical structure of the
target space it is useful to formulate the $\mathcal{N}$ = (2,2)
non-linear $\sigma$-model in terms of an $\mathcal{N}$ = (2,2)
superspace. The most general Lagrangian density one can write down
in $\mathcal{N}$ = (2,2) superspace consists of (the superspace
integration) of a real scalar potential, as can be seen from
dimensional analysis. This scalar potential is a function of three
types of $\mathcal{N}$ = (2,2) scalar superfield (chiral, twisted
chiral and semi-chiral) and is naturally interpreted as the
generalized K\"{a}hler potential, as we will
explain in the first part of section \ref{sec2}.\\
\indent It is well known that the spectrum of type II superstrings
also contains open string states describing the excitations of
D-branes to which the open string is attached. Therefore, it is a
cromulent question whether we can use these two-dimensional
$\mathcal{N}$ = (2,2) $\sigma$-models to describe the propagation of
an open string in (generalized) K\"{a}hler backgrounds with
(supersymmetric) D-branes. We can invoke the effects of a
(supersymmetric) D-brane quite easily by noticing that the boundary
conditions of the open string will break half of the supersymmetry
to an $\mathcal{N}$ = 2 boundary supersymmetry. To have a manifestly
supersymmetric description of the boundary conditions for an open
string we need to introduce an $\mathcal{N}$ = 2 boundary
superspace. The picture for exclusively chiral or exclusively
twisted chiral superfields was developed in \cite{Sevrin:2007yn},
which lead to type B and respectively type A branes on K\"{a}hler
target spaces. The picture for chiral and twisted chiral superfields
was successfully unraveled in \cite{Sevrin:2008} and is reviewed in
\cite{SSW:2009pr1}. A brief review of the picture will be given in
section \ref{sec2}, followed by some examples of D-branes wrapped on
(generalized) K\"{a}hler geometries.
\\
\indent In the third section we start by discussing the main
philosophy of T-dualization in the $\mathcal{N}$ = (2,2) superspace
formalism, which basically corresponds to a Legendre transformation
interchanging chiral and twisted chiral superfields. The natural
follow-up question is whether we can perform duality transformations
in the presence of boundaries. We will point out that it is
possible, though there are some subtleties to take into account.
Knowing these subtleties allowed us to come up with a systematic
procedure to construct coisotropic D-branes via T-dualization of a
chiral superfield \cite{Sevrin:2008}. We will conclude section
\ref{sec3} with some explicit examples.

\section{The boundary superspace approach and D-branes}
\label{sec2}
Let us start by giving a quick review of the $\mathcal{N}$ = (2,2)
non-linear $\sigma$-model in an $\mathcal{N}$ = (2,2) superspace.
The world-sheet bosonic coordinates are given by ($\sigma$, $\tau$),
while we also introduce four Grassmann coordinates ($\theta^+$,
$\theta^-$, $\hat \theta^+$, $\hat \theta^-$) and associated
super-covariant derivatives\footnote{Note that we use a different
basis w.r.t. \cite{SSW:2009pr1}. For conventions we refer to
\cite{Sevrin:2007yn, Sevrin:2008}.} ($\ID_+$, $\ID_-$, $\bar \ID_+$,
$\bar \ID_-$). The most general action in $\mathcal{N}$ = (2,2)
superspace reads on dimensional grounds,
\begin{eqnarray}
{\cal S}=4\,\int\,d^2 \sigma \,d^2\theta \,d^2 \hat \theta \, V(X,
\bar X), \label{1eq1}
\end{eqnarray}
where the lagrange density $V(X, \bar X)$ is a real function of
complex $\mathcal{N}$ = (2,2) scalar superfields. A closer look at
the number of degrees of freedom of the $\mathcal{N}$ = (2,2)
superfields indicates that there are too many degrees of freedom in
comparison with an $\mathcal{N}$ = (2,2) $\sigma$-model in
$\mathcal{N}$ = (1,1) superspace. In order to eliminate some of the
degrees of freedom we should impose constraints on the $\mathcal{N}$
= (2,2) scalar superfields. We can distinguish three different types
of constraint, leading to three different types of superfield:
chiral, twisted chiral and semi-chiral superfields. However, in the
remainder of the note we shall mostly focus on the first two types:\\
$\bullet$ chiral superfields $z^\alpha$, $z^{\bar \alpha}$ with
$\alpha, \bar\alpha \in \{1,\cdots,n_c\}$
\begin{eqnarray}
 \bar \ID_{\pm} z^\alpha = 0,\qquad  \ID_\pm z^{ \bar \alpha }=0,
\label{1eq2}
\end{eqnarray}
$\bullet$ twisted chiral superfields $w^\mu$, $w^{\bar \mu}$ with
$\mu,\bar \mu \in \{1,\cdots,n_t \}$
\begin{eqnarray}
\bar \ID_+ w^ \mu = 0 = \ID_- w^\mu,\qquad \ID_+ w^{ \bar \mu }= 0 =
\bar \ID_- w^{ \bar \mu }, \label{1eq3}
\end{eqnarray}
$\bullet$ semi-chiral superfields $l^{\tilde\alpha},
l^{\bar{\tilde\alpha}}, r^{\tilde \mu}, r^{\bar{ \tilde \mu}} $ with
$\tilde\alpha, \bar{\tilde\alpha}, \tilde\mu, \bar{\tilde\mu} \in
\{1,\cdots,n_s\} $
\begin{eqnarray}
\bar \ID_+ l^{\tilde\alpha} = 0 , \qquad \ID_+ l^{\bar{\tilde
\alpha}} = 0, \qquad \bar \ID_- r^{\tilde \mu} = 0, \qquad \ID_-
r^{\bar{\tilde \mu}} = 0 \label{1eq4}
\end{eqnarray}
Introducing these constraints makes it also possible to investigate
the dynamics of the $\mathcal{N}$ = (2,2) $\sigma$-models, which is
clearly not present in action eq.~(\ref{1eq1}). The proper way to
see the dynamics is by reducing the action eq.~(\ref{1eq1}) to
$\mathcal{N}$ = (1,1) superspace by integrating out $\bar \ID_+$ and
$\bar \ID_-$. Comparing the resulting action with the most general
$\mathcal{N}$ = (1,1) superspace action\footnote{The most general
$\mathcal{N}$ = (1,1) superspace action can e.g. be found in
\cite{SSW:2009pr1}.}, we can read off the expressions for the metric
$g_{ab}$ and the 2-form potential $b_{ab}$ in terms of $V$. In case
of chiral and twisted chiral fields, we get the following
expressions,
\begin{eqnarray}
&&g_{ \alpha \bar \beta }=+V_{ \alpha \bar \beta },\qquad g_{ \mu
\bar \nu }= -
V_{ \mu \bar \nu },\nonumber\\
&&b_{ \alpha \bar \nu }=-V_{ \alpha \bar \nu }, \qquad b_{ \mu \bar
\beta  }=+V_{ \mu \bar \beta  }, \label{1eq5}
\end{eqnarray}
and all other components\footnote{An expression of the form $V_{
\alpha \bar \nu }$ is a shorthanded notation for $\partial_\alpha
\partial_{\bar \nu} V$, etc.} vanish.  From these relations it is
obvious that $V$ plays the role of the (generalized) K\"{a}hler
potential.
\\
\indent Let us now give some examples of four-dimensional target
spaces that can be parameterized by chiral and/or twisted chiral
superfields. The easier target spaces are the torus $T^4$ and
$D\times T^2$, where $D$ represents the disk with a singular
boundary. Both target spaces can be parameterized by two chiral
superfields, two twisted chiral superfields, or one chiral and one
twisted chiral superfield. A third example is the Wess-Zumino-Witten
(WZW) model on $SU(2)\times U(1)$ \footnote{$SU(2)\times U(1)$ can
also be seen as the Hopf surface $S^3\times S^1$.}, which is the
only $\mathcal{N}$ = (2,2) WZW model that can be parameterized
without the use of semi-chiral superfields \cite{Rocek:1991vk}, in which case it is
parameterized by one chiral and one twisted chiral superfield. The target space is
characterized by a metric and a torsion. This concludes our
discussion of a closed string propagating on (generalized)
K\"{a}hler backgrounds.
\\
\indent Next, we will discuss the propagation of an open string in
the presence of a D-brane to which the open string is attached.
Invoking the presence of a D-brane comes down to introducing a
boundary that breaks the re-parametrization invariance along the
open string and half of the supersymmetries on the world-sheet. In
practice it is sufficient to recombine the supercovariant
derivatives ($\ID_\pm, \bar \ID_\pm$) into the following linear
combinations,
\begin{eqnarray}
\ID\equiv \ID_+ + \ID_-,\qquad \bar \ID\equiv \bar \ID_+ + \bar
\ID_-, \qquad \ID'\equiv \ID_+ - \ID_-,\qquad \bar \ID'\equiv \bar
\ID_+ - \bar \ID_-,\label{1eq6}
\end{eqnarray}
where $\ID$ and $\bar \ID$ represent the directions which remain
invariant.
In the next step we rewrite the superfield constraints
eqs.~(\ref{1eq2}) and (\ref{1eq3}) in terms of these new
supercovariant derivatives to arrive at $\mathcal{N}$ = 2 boundary
superfields:\\
$\bullet$ chiral boundary superfields $z^\alpha$, $z^{\bar \alpha}$
\begin{eqnarray}
\bar \ID z^ \alpha= 0 = \bar \ID'z^ \alpha, \qquad \ID z^{ \bar
\alpha } = 0 = \ID' z^{ \bar \alpha },\label{1eq7}
\end{eqnarray}
$\bullet$ twisted chiral boundary superfields $w^\mu$, $w^{\bar
\mu}$
\begin{eqnarray}
\ID'w^ \mu = \ID w^ \mu , \, \bar \ID'w^ \mu =- \bar \ID w^ \mu ,
\qquad \ID'w^{ \bar \mu } =-\ID w^ { \bar \mu }, \, \bar \ID'w^{
\bar \mu } = \bar \ID w^ { \bar \mu }. \label{1eq8}
\end{eqnarray}
These manipulations form the basic procedure to arrive at the
$\mathcal{N}$ = 2 boundary superspace formalism. Finally we write
down the most general $\mathcal{N}$ = 2 boundary superspace action
with chiral and twisted chiral superfields as,
\begin{eqnarray}
{\cal S}=- \frac{1}{4}\int d^2 \sigma\, d \theta d \hat \theta\,
\ID' \bar \ID'\, V(z^\alpha, w^\mu, z^{\bar\alpha}, w^{\bar\mu} )+
i\,\int d \tau \,d \theta d \hat \theta \,W( z^\alpha, w^\mu,
z^{\bar\alpha}, w^{\bar\mu}),\label{1eq9}
\end{eqnarray}
where $V$ is the (real) bulk potential and $W$ the (real) boundary
potential, and where we have to integrate out $\ID'$ and $\bar
\ID'$.
We can describe the local embedding of a D-brane by examining the
boundary conditions of the open string attached to the D-brane. We
therefore vary the action eq.~(\ref{1eq9}) with respect to the
chiral and twisted chiral superfields, after integrating out the
derivatives $\ID'$ and $\bar \ID'$. This variation will yield a bulk
term and a boundary term. The bulk term will describe the
propagation of the bulk of the open string and contains information
about the target space geometry on which the string is propagating.
The boundary term describes the propagation of the endpoints of the
open string and thus contains information about the local D-brane
geometry. For a complete analysis we refer to \cite{Sevrin:2007yn,
Sevrin:2008}. In this note we shall limit ourselves to
four-dimensional target spaces, parameterized by chiral and/or
twisted chiral superfields, to see how the analysis works in practice.\\
\indent In order for the boundary variation to vanish it is
necessary to impose appropiate boundary conditions. These boundary
conditions are determined by the bulk potential $V$, the boundary
potential $W$ and the constraints on the boundary superfields. For
the chiral superfield the boundary variation allows us to impose a
Dirichlet or a Neumann boundary condition, but the boundary
superfield constraints imply the same type of boundary condition for
the superfield and its complex conjugated. Thus, we can impose
two Dirichlet boundary conditions or two Neumann boundary conditions
for one chiral boundary superfield. This means that we can have
three different types of branes (so-called B-branes) for a
four-dimensional target space parameterized by purely chiral
superfields: D0-, D2- and D4-branes. The boundary constraints for a
twisted chiral superfield imply that every Dirichlet condition
should be accompanied by an associated Neumann boundary condition.
However, in the case of two (or more) twisted chiral superfields,
one can impose four independent Neumann boundary conditions. Thus,
on a four-dimensional target space parameterized by purely twisted
chiral superfields we can wrap two different types of brane
(so-called A-branes): lagrangian
D2$_\ell$-branes and (space-filling) coisotropic D4$_c$-branes.\\
\indent In the case of a four-dimensional target space parameterized
by a chiral and a twisted chiral superfield, we will always have one
Dirichlet and one Neumann boundary condition for the twisted chiral
field and two Dirichlet or two Neumann boundary conditions for the
chiral field. Hence, we encounter two different types of brane:
D1-branes and D3-branes. Now, we are able to give an overview table
with the different possible D-branes wrapping a subspace of a
(generalized) K\"{a}hler geometry and preserving half of the
$\mathcal{N}$ = (2,2) world-sheet supersymmetry,
\begin{table}[h]
\begin{center}
\begin{tabular}{@{}ccc@{}}
\hline field content & geometry & branes \\
\hline 2 chiral & K\"{a}hler & D0, D2, D4\\
1 chiral + 1 twisted chiral & (Generalized) K\"{a}hler & D1, D3 \\
2 twisted chiral & K\"{a}hler & D2$_{\ell}$, D4$_{c}$ \\
\hline
\end{tabular}
\caption{Possible D-brane configurations for a four-dimensional
target space} \label{1tab1}
\end{center}
\end{table}
\\
\indent In \cite{Sevrin:2008} we constructed a D3-brane on $T^4$ and
$S^3\times S^1$. The Dirichlet boundary condition
of the D3-brane on $T^4$ is given by,
\begin{eqnarray}
 \alpha \,w+\bar \alpha \,\bar w=\beta \,z+\bar \beta \,\bar z,\label{1eq10}
\end{eqnarray}
where $\alpha ,\,\beta \in\IZ+i\,\IZ$ and $\alpha \neq 0$.  In the
example of $S^3\times S^1$ we choose the following Dirichlet
boundary condition\footnote{
One can make this embedding less mysterious by introducing the Hopf
coordinates $z=\cos \psi \,e^{ \rho +i \phi _1}$, $w=\sin \psi \,e^{
\rho + i\phi _2}$, with $ \phi _1,\, \phi_2,\, \rho \in
\IR\,\mbox{mod}\,2 \pi  $ and $\psi\in[0, \pi /2]$.},
\begin{eqnarray}
-i \ln \frac{w}{ \bar w}=m_1\,x+m_2\,y,\label{1eq11}
\end{eqnarray}
where $m_1$, $m_2$ $\in \IZ$. For simplicity we defined $x \equiv
\ln ( z\bar z+w\bar w)$ and $y \equiv -i\,\ln\frac{ z}{ \bar z}$.
Both types of D3-brane are non-trivial embeddings satisfying all
consistency requirements and made possible due to the presence of a
$U(1)$ gauge field on the D3-brane world-volume.

\section{Duality transformations in extended superspace}\label{sec3}
T-duality is another important feature of type II superstring models
and it is therefore interesting to see how T-duality can be made
manifest in the $\mathcal{N}$ = (2,2) non-linear $\sigma$-models.
Like in the previous section we prefer to work in $\mathcal{N}$ =
(2,2) superspace, where a T-duality corresponds to a
Legendre-transformation interchanging different types of
$\mathcal{N}$ = (2,2) superfield \cite{Gates:1984nk, GMST:98}.\\
\indent We will first summarize the basic procedure to dualize a
chiral or a twisted chiral superfield in the absence of boundaries.
In order to dualize on the level of the action we have to assume
that the model in eq.~(\ref{1eq1}) exhibits an isometry of the form
$X + \bar X$. This isometry then needs to be gauged on the
world-sheet, which demands the introduction of a real $\mathcal{N}$
= (2,2) gauge superfield $Y$ to preserve the (local) isometry. Since
the gauge field $Y$ should not introduce extra degrees of freedom,
we should impose that $Y$ is purely gauge. This can be done through
a complex $\mathcal{N}$ = (2,2) superfield (serving as a
Lagrange-multiplier) imposing that the field strengths derived from
the gauge field $Y$ vanish. The gauged K\"ahler potential together
with the Lagrange multiplier terms form the complete first order
potential. If we integrate out the Lagrange multipliers from this
first order action, we retrieve the original model. To arrive at the
dual model we need to integrate out the gauge field and the dual
superfield $\tilde X$ will be expressed in terms of superspace
derivates of the Lagrange-multiplier. Performing this philosophy we
can dualize a chiral field to a twisted chiral field and vice versa.
Let us conclude the discussion of T-dualization in the absence of
boundaries by giving some concrete four-dimensional examples. The
torus $T^4$ parameterized by a chiral and a twisted chiral
superfield can be dualized to a dual torus $T^4$ parameterized by
two chiral superfields or two twisted chiral superfields. The WZW
model on the Hopf-surface $S^3\times S^1$ parameterized by a chiral
and a twisted chiral superfield can be dualized to $D\times T^2$
parameterized by two chiral superfields or two twisted chiral
superfields.
\\
\indent The next step is to translate this general philosophy to
$\mathcal{N}$ = 2 boundary superspace and to investigate which kind
of D-brane configuration we get after T-dualization. An initial
analysis of duality transformations in the presence of boundaries
can be found in \cite{Sevrin:2007yn}. There it was already realized
that one should pay extra attention to the boundary term in order to
get consistent dual boundary conditions. First of all we might want
to rewrite the boundary potential such that the symmetry of the form
$X + \bar X$ remains present at the boundary, albeit not necessarily
explicitly. To arrive at the correct and consistent boundary
conditions, we might also need to add extra terms to the boundary
action. This enabled us for instance to dualize a space-filling
B-brane on a K\"ahler target space to a lagrangian A-brane on the
dual K\"ahler target space, and a lagrangian D1-brane on $T^2$ to a
space-filling D2-brane on the dual $T^2$. Moreover, starting from a
coisotropic D4-brane on a four-dimensional hyper-K\"ahler target
space we were able to construct a D3-brane on a generalized K\"ahler
target space via dualization. In \cite{Sevrin:2008} we improved the
dualization method in the presence of boundaries by deriving two
identities eqs.~(6.14) and (6.15) that made it possible to introduce
the correct boundary terms. With these two identities all
dualizations interchanging chiral and twisted chiral superfields can
be performed. \\
\indent We will not dwell too long on the general philosophy of
T-dualization in superspace and try to clarify it by reviewing some
practical examples. We shall focus again on the D3-branes
constructed on $T^4$ and $S^3\times S^1$, as discussed in section
\ref{sec2}. It will become clear through these examples that the
parameters describing the D3-brane embedding determine the
characteristics of the dual brane. Let us start by dualizing the
chiral superfield. Looking at the Dirichlet boundary conditions
eqs.~(\ref{1eq10}) and (\ref{1eq11}) we see that the Dirichlet
boundary condition preserves the symmetry of the form $z + \bar z$
if $\beta = \bar \beta$, and $m_2 = 0$ respectively and that one of
the D3-brane directions is wrapped along the $z + \bar z$ direction.
When dualizing along this direction we expect a (lagrangian)
D2-brane in the dual theory. For the D3-brane on $S^3 \times S^1$ we
find as the dual model a lagrangian D2-brane wrapped along one
direction in $D$ and one direction in $T^2$, where the (quantized)
wrapping angle is given by $m_1$. On the other hand, if $\beta \neq
\bar \beta$, and $m_2 \neq 0$ respectively, then the Dirichlet
boundary conditions eqs.~(\ref{1eq10}) and (\ref{1eq11}) violate the
symmetry of the form $z + \bar z$. The D3-brane is wrapped
differently and when dualizing along the direction $z + \bar z$ we
expect a D4-brane in the dual theory. In the case of the (dual)
$T^4$ the constructed space-filling coisotropic D4-brane is a
generalization of the coisotropic D4-brane described in
\cite{Sevrin:2007yn}. The coisotropic D4-brane on $D\times T^2$ is a
quite interesting result, since it is a first example of a
coisotropic D-brane on a non-hyper-K\"{a}hler target space. From a
target space perspective the dual target space should allow for a
(second) complex structure $K$, which does not commute with the
complex structures $J_{(\pm)}$ characterizing the target space
geometry. On the world-volume of the coisotropic D4-brane lives a
$U(1)$ fieldstrength that can be given in terms of the complex
structures and the metric (see e.g. eq.~(4.61) in
\cite{Sevrin:2008}).
\\
\indent In our case, the models also exhibit the isometry $w + \bar
w$ in the bulk for the twisted chiral superfield. We were able to
dualize the D3-branes along this isometry direction, leaving us with
dual models completely parameterized by two chiral superfields. In
the case of the D3-brane on $T^4$ we can distinguish once more two
different cases, i.e. $\alpha = \bar \alpha$ and $\alpha \neq \bar
\alpha$. In the first case the D3-brane is wrapped along the
direction we dualize (i.e. $w + \bar w$) and so we expect a D2-brane
in the dual model. The D2-brane is now wrapping a holomorphic
2-cycle with a non-trivial $U(1)$ bundle on its world-volume. The
dualization on the level of the action is rather subtle for this
case and for details we refer to \cite{Sevrin:2008}. In the latter
case the D-brane is no longer wrapped along the dualization
direction and we find a D4-brane wrapping a holomorphic 4-cycle with
a non-trivial $U(1)$ bundle on its world-volume. The constructed
D3-brane on $S^3 \times S^1$ can only be dualized to a D4-brane
wrapping a holomorphic 4-cycle on $D\times T^2$ with a non-trivial
$U(1)$-bundle on its world-volume.\\
\\
\indent Besides the dualization of only one type of superfield, we
would also like to briefly consider here the dualization of a chiral
and twisted chiral pair to a semi-chiral supermultiplet
\cite{GMST:98}. The underlying gauge structure and T-duality
transformations were discussed in \cite{LRRUZ:07, GM:07}. In
\cite{Sevrin:2008} we started the analysis of dualizing a chiral and
twisted chiral pair to a semi-chiral supermultiplet in the presence
of boundaries. Starting from the torus $T^4$ parameterized by a
chiral/twisted chiral pair and the D3-brane given in
eq.~(\ref{1eq10}), we were able to dualize the D3-brane to a
lagrangian-like D2-brane and a coisotropic-like D4-brane, depending
on the embedding parameters of the D3-brane. This dualization
allowed us to have a quick look at the possible boundary conditions
for semi-chiral superfields and initiated the study of semi-chiral
boundary superfields, which will be continued in \cite{SSW:2008}. In
this upcoming paper, we will also consider the dualization of
$D\times T^2$ parameterized by a chiral/twisted chiral pair to
$S^3\times S^1$ parameterized by a semi-chiral multiplet. Through
this dualization it is possible to construct a lagrangian-like
D2-brane and a coisotropic-like D4-brane on $S^3\times S^1$,
starting from a D3-brane on $D\times T 2$.

\section{Closing remarks and outlook}
In this note we briefly discussed a manifestly $\mathcal{N}$ = 2
supersymmetric world-sheet description of D-branes wrapping
subspaces of bihermitian geometries with commuting complex
structures. For this type of geometries, the target space is
parameterized by a set of chiral and/or twisted chiral superfields.
We gave some four-dimensional examples parameterized by a chiral and
a twisted chiral superfield to clarify the general formalism
developed in \cite{Sevrin:2008} and discussed in \cite{SSW:2009pr1}.
Building on the results of \cite{Sevrin:2007yn} we explored the
subtleties in dualizing one of the superfields when boundaries are
present. This exploration led to a systematic method to construct
coisotropic D-branes via dualization of the chiral superfield, which
allowed us to construct a coisotropic D4-brane on a
non-hyper-K\"{a}hler target space. In \cite{Sevrin:2008} we also
dualized the chiral/twisted chiral pair to one semi-chiral
multiplet, which gave a brief taste of the $\mathcal{N}$ = 2
boundary superspace description of D-branes on generalized
K\"{a}hler manifolds parameterized by semi-chiral superfields. This
will be discussed in depth in \cite{SSW:2008} and reviewed in
\cite{SSW:2009pr1}.\\
\indent In a later stage, it would be interesting to check the
stability of the constructed D-branes by studying the quantum
conformal invariance of the two-dimensional models using a
$\mathcal{N}$ = 2 boundary superspace approach, as was done in
\cite{NSTW:2006}. Studying the quantum conformal invariance of these
models will also allow us to investigate the number of target space
supersymmetries preserved by the constructed D-branes.

\begin{acknowledgement}
We thank Ulf Lindstr\"{o}m, Martin Ro\v{c}ek and Maxim Zabine for
useful discussions and suggestions. W.S. and A.W. would like to thank 
the organizers of the fourth EU RTN workshop in Varna for the opportunity to present their work. 
All authors are supported in part by the European Commission FP6 RTN programme
MRTN-CT-2004-005104. AS and WS are supported in part by the Belgian
Federal Science Policy Office through the Interuniversity Attraction
Pole P6/11,  and in part by the ``FWO-Vlaanderen'' through project
G.0428.06. AW is supported in part by grant 070034022 from the
Icelandic Research Fund.
\end{acknowledgement}



\begin{thebibliography}{99}
\bibitem{Zumino:1979}
  B.~Zumino,
  Phys.~Lett.~{\bf B 87}~(1979)~{203}.
\bibitem{Alvarez-Gaume:1981hm}
  L.~Alvarez-Gaume and D.~Z.~Freedman,
  Commun.~Math.~Phys.~{\bf 80}~(1981)~{443}.
\bibitem{Gates:1984nk}
  S.~J.~Gates, C.~M.~Hull and M.~Ro\v cek,
  Nucl.~Phys.~{\bf B 248}~(1984)~{157}.
\bibitem{Buscher:1987uw}
  T.~Buscher, U.~Lindstr{\"o}m and M.~Ro\v cek,
  Phys.~Lett.~{\bf B 202}~(1988)~{94}.
\bibitem{Sevrin:1996jr}
  A.~Sevrin and J.~Troost,
  Nucl.\ Phys.\ {\bf B 492 } (1997) {623}, {\tt [arXiv:hep-th/9610102]}.
\bibitem{Lindstrom:2005zr}
  U.~Lindstrom, M.~Rocek, R.~von Unge and M.~Zabzine,
  Commun.~Math.~Phys.~{\bf 269}~(2007)~{833},
  [arXiv:hep-th/0512164].
\bibitem{Gualtieri:2007ng}
  M.~Gualtieri,
{\tt arXiv:math/0401221, arXiv:math/0703298}.
\bibitem{Sevrin:2007yn}
  A.~Sevrin, W.~Staessens and A.~Wijns,
  JHEP~{\bf 0711}~(2007)~{061},
  {\tt arXiv:0709.3733 [hep-th]}.
\bibitem{Sevrin:2008}
  A.~Sevrin, W.~Staessens and A.~Wijns,
  JHEP~{\bf 0810}~(2008)~{108}
  {\tt arXiv:0809.3659 [hep-th]}.
\bibitem{SSW:2009pr1}
  A.~Sevrin, W.~Staessens and A.~Wijns,
  {\tt arXiv:0810.5355 [hep-th]}
\bibitem{Rocek:1991vk}
  M.~Ro\v cek, K.~Schoutens and A.~Sevrin,
  Phys.~Lett.~{\bf B 265} (1991) 303.
\bibitem{SSW:2008}
  A.~Sevrin, W.~Staessens and A.~Wijns,
  work in progress.
\bibitem{GMST:98}
 M.~T.~Grisaru, M.~Massar, A.~Sevrin, J.~Troost,
 {\tt [arXiv:hep-th/9801080]}.
\bibitem{LRRUZ:07}
 U.~Lindstr\"om, M.~Ro\v{c}ek, I.~Ryb, R.~von Unge, and M.~Zabzine,
 {\tt arXiv:0705.3201 [hep-th]}.
\bibitem{GM:07}
 S.~J.~Gates~Jr., W.~Merrell,
 {\tt arXiv:0705.3207 [hep-th]}.
\bibitem{NSTW:2006}
  S.~Nevens, A.~Sevrin, W.~Troost, A.~Wijns,
  JHEP~{\bf 0608}~(2006)~{086}, {\tt arXiv:hep-th/0606255}.
\end{thebibliography}
\end{document}